\documentclass[a4paper]{jpconf}
\usepackage{graphicx}
\begin{document}
\title{Nonadditive entropy and nonextensive statistical mechanics - Some central concepts and 
recent applications}
 
\author{Constantino Tsallis$^{1,2}$ and Ugur Tirnakli$^{3,4}$}

\address{$^1$ Centro Brasileiro de Pesquisas Fisicas and National Institute of Science and Technology 
for Complex Systems, Rua Xavier Sigaud 150, 22290-180 Rio de Janeiro-RJ, Brazil\\
$^2$ Santa Fe Institute, 1399 Hyde Park Road, Santa Fe, NM 87501, USA\\
$^3$ Department of Physics, Faculty of Science, Ege University, 35100 Izmir, Turkey\\
$^4$ Division of Statistical Mechanics and Complexity, 
Institute of Theoretical and Applied Physics (ITAP) Kaygiseki Mevkii, 
48740 Turunc, Mugla, Turkey}

\ead{tsallis@cbpf.br, ugur.tirnakli@ege.edu.tr}

\begin{abstract}
We briefly review central concepts concerning nonextensive statistical mechanics, based on the 
nonadditive entropy $S_q=k\frac{1-\sum_{i}p_i^q}{q-1}\;(q \in {\cal R}; \,S_1=-k\sum_{i}p_i \ln p_i)$. 
Among others, we focus on possible realizations of the $q$-generalized Central Limit Theorem, including 
at the edge of chaos of the logistic map, and for quasi-stationary states of many-body long-range-interacting 
Hamiltonian systems.
\end{abstract}

\section{Introduction}

\subsection{Entropy}

The ubiquitous concept of {\it energy} is associated with the {\it possibilities} for the configurations of 
a mechanical system (e.g., the eigenvalues of the Hamiltonian of a quantum system defined in a specific 
Hilbert space). The concept of {\it entropy} emerges in an even larger domain, since it can be defined for 
any system, mechanical or not, which admits a set of {\it probabilities} for its possible configurations. 
For instance, if we are dealing with a quantum mechanical system, the set of probabilities typically is 
that corresponding to the eigenvectors of the Hilbert space. Epistemologically speaking, entropy is 
one of the most subtle concepts in physics. Entropy and energy together constitute the basis on which statistical mechanics --- one of the pillars of contemporary physics --- is constructed.

The entropy, initially defined by Clausius for thermodynamics, connects the macroscopic and microscopic 
worlds. Its most elementary form is the logarithmic one, first introduced by Boltzmann and refined by Gibbs, 
von Neumann, Shannon, Jaynes and others. For a finite discrete set of probabilities $\{p_i\}$ is given by
\begin{equation}
S_{BG}=-k\sum_{i=1}^Wp_i \ln p_i \;\;\;\;\left(\sum_{i=1}^Wp_i=1\right)\,,
\label{BGentropy}
\end{equation}
where $BG$ stands for {\it Boltzmann-Gibbs}. The conventional constant $k$ is typically taken to be the 
Boltzmann universal constant for thermostatistical systems, or taken to be unity in information 
theory. 
For the particular case of equal probabilities, i.e., $p_i=1/W , \,\forall i$, we have 
\begin{equation}
S_{BG}=k \ln W \,,
\end{equation}
carved on stone in Boltzmann grave in Vienna. Expression (1) enables the construction of 
a remarkably useful physical theory, referred to as {\it Boltzmann-Gibbs statistical mechanics}.

Many entropic forms have been introduced, and reintroduced, since Boltzmann (see 
\cite{Tsallis2009a,Tsallis2009b} and references therein for details). In 1988 \cite{Tsallis1988}, a more 
general form, namely
\begin{equation}
S_{q}=k\frac{1-\sum_{i=1}^Wp_i^q}{q-1} \;\;\;\;\left(\sum_{i=1}^Wp_i=1; \,q\in {\cal R}; \,S_1=S_{BG}\right)\,,
\label{qentropy}
\end{equation}
was proposed as the basis for generalizing $BG$ statistical mechanics (into a theory now known as 
{\it nonextensive statistical mechanics} \footnote{The word {\it nonextensive} is to be associated with the 
fact that the total energy of long-range-interacting mechanical systems is nonextensive, in contrast with the 
case of short-range-interacting systems, whose total energy is {\it extensive} in the thermodynamical sense.}). For the particular case of equal probabilities, expression (\ref{qentropy}) becomes 
\begin{equation}
S_{q}=k \frac{W^{1-q}-1}{1-q} \,.
\label{qentropyequal}
\end{equation}

By introducing the $q-logarithmic$ function
\begin{equation}
\ln_q x \equiv \frac{x^{1-q}-1}{1-q} \;\;\;(x>0; \,q \in {\cal R};\,\ln_1 x=\ln x)\,,
\label{qlogarithmic}
\end{equation}
expressions (\ref{qentropy}) and (\ref{qentropyequal}) can be respectively rewritten as follows:
\begin{equation}
S_{q}=k\sum_{i=1}^Wp_i \ln_q \frac{1}{p_i}=-k\sum_{i=1}^Wp_i^q \ln_q p_i\,,
\label{BGentropy}
\end{equation}
and
\begin{equation}
S_q=k\ln_qW\,.
\end{equation}

The axiomatics associated with this entropy have been quite explored and interesting characterizations have 
emerged. For details we may address the reader to \cite{Santos1997,Abe2000,Topsoe2006,Topsoe2010,OharaMatsuzoeAmari2010}, 
among others.

\subsection{Additivity versus extensivity}

We adopt for {\it entropy additivity} the definition given in Penrose's classical book \cite{Penrose1970}, 
namely that an entropy $S$ is said {\it additive} if, for any two {\it probabilistically independent} systems 
$A$ and $B$, i.e., for $p_{i,j}^{A+B}=p_i^A p_j^B, \,\forall (i,j)$, we have that 
\begin{equation}
S(A+B)=S(A)+S(B)\,,
\end{equation}
where $S(A+B)\equiv S(\{p_{i,j}^{A+B}\})$, $S(A)\equiv S(\{ p_i^A\})$, and $S(B)\equiv S(\{ p_i^B\})$. 

>From definition (\ref{qentropy}) it is straightforward to prove that, for any two probabilistically 
independent systems $A$ and $B$,
\begin{equation}
\frac{S_q(A+B)}{k}=\frac{S_q(A)}{k}+\frac{S_q(B)}{k}+(1-q)\frac{S_q(A)}{k}\frac{S_q(B)}{k} \,.
\end{equation}
Therefore, $S_{BG}$ is additive, whereas $S_q$ $(q \ne 1)$ is nonadditive \footnote{We remark that additivity 
is obtained whenever $(1-q)/k \to 0$. We see that this can occur in two different manners: $q\to 1$ for fixed 
$k$, and $k \to \infty$ for fixed $q$. The latter corresponds to the infinite temperature $T$ limit, since in 
all thermostatistical systems $T$ always appears in the form $kT$, thus having the dimension of an energy.}.

{\it Entropic extensivity} is a concept in some sense more subtle than additivity. An entropy $S$ of a given 
system constituted by $N$ elements is said to be {\it extensive} if
\begin{equation}
0< \lim_{N\to\infty}\frac{S(N)}{N}<\infty \,,
\end{equation}
i.e., if $S(N) \propto N $ for $N>>1$. We see therefore that additivity only depends on the specific 
mathematical connection between the macroscopic entropy functional and the probabilities of the configurations 
of the system. Extensivity depends on this but {\it also} on the specific system, more precisely on the nature 
of the correlations of its elements, and therefore of its collective configurations.

The distinction between additivity and extensivity has already been illustrated in simple probabilistic 
systems \cite{TsallisGellMannSato2005}. It has also been shown for the so-called {\it block entropy} 
(entropy of a subsystem of the entire system) of strongly quantum entangled fermionic and bosonic systems 
\cite{CarusoTsallis2008}. 

For the probabilistic system it has been shown \cite{TsallisGellMannSato2005} that 
\begin{equation}
S_{q_{ent}}(N) \propto N \;\;\;(N>>1)\,,
\end{equation} 
where $ent$ stands for {\it entropy}, $N$ is the number of (strongly correlated) binary random variables, and 
\begin{equation}
q_{ent}=1-\frac{1}{d}\,,
\end{equation} 
$d=1,2,3,...$ characterizing the width of an infinitely long strip of nonvanishing probabilities of a 
probability triangle asymptotically satisfying the Leibnitz rule (i.e., asymptotically scale-invariant).

For the (one-dimensional) fermionic system it has been shown that, at criticality at vanishing temperature, 
we have \cite{CarusoTsallis2008}
\begin{equation}
\lim_{N\to\infty}S_{q_{ent}}(N,L) \propto L \;\;\;(L>>1)\,,
\end{equation}
where $L$ is the number of first-neighboring spins (or analogous elements) within an infinitely long ($N\to\infty$) chain, and
\begin{equation}
q_{ent}=\frac{\sqrt{9+c^2}-3}{c}\,, 
\end{equation} 
where $c$ is the {\it central charge} ($c=1/2$ for the Ising ferromagnet, and $c=1$ for the isotropic $XY$ 
ferromagnet, in the presence of a critical transverse magnetic field in both cases). We verify that $q_{ent}$ 
monotonically increases from zero to one when $c$ increases from zero to infinity.

For the (two-dimensional) bosonic system the results are qualitatively the same. However they have been 
established only numerically, not analytically.

\begin{figure}[h]
\includegraphics[width=28pc]{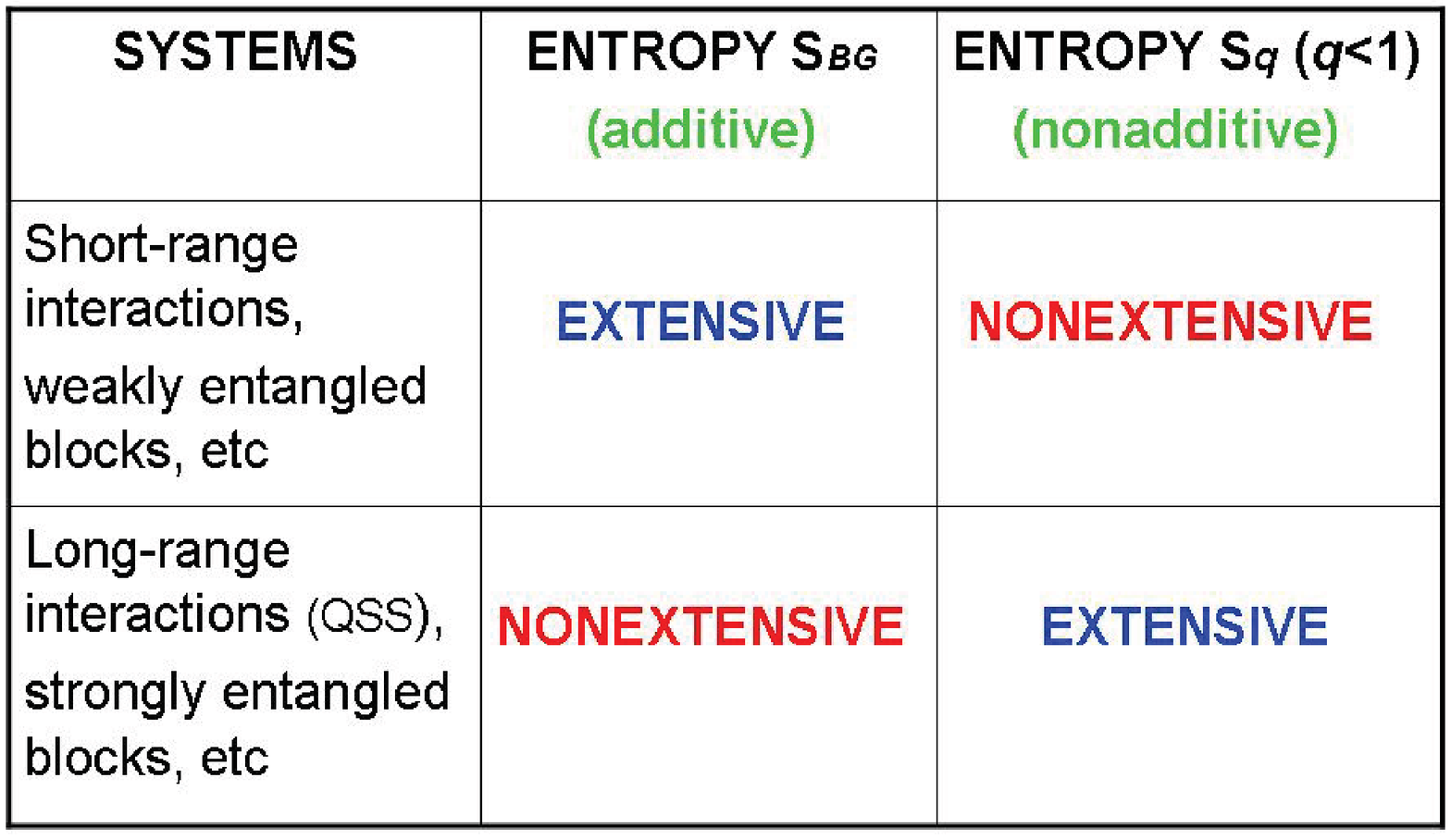}
\caption{Entropic additivity and entropic extensivity are different concepts. The BG entropy is additive, 
whereas $S_q$ (for $ q\ne 1$) is nonadditive. Extensivity depends on the system: for standard systems, 
$S_{BG}$ is extensive, whereas it can be nonextensive for anomalous systems; $S_q$ is the other way around.}
\label{extensive}
\end{figure}

The generic scenario which emerges is that, for a vast class of systems (but certainly not all), a value 
$q_{ent}$ exists such that $S_{q_{ent}}(N) \propto N\;\;(N \to\infty)$, where $N$ is the number of elements 
of the system under consideration (which might be the entire system, or only a large part of it). For standard 
systems, we have that $q_{ent}=1$; for various classes of anomalous systems, we have $q_{ent} \ne 1$. 
The situation is depicted in Fig. \ref{extensive}.

\subsection{Central limit theorems}

\begin{figure*}[ht!]
\begin{center}
\includegraphics[width=15.0 cm,angle=0]{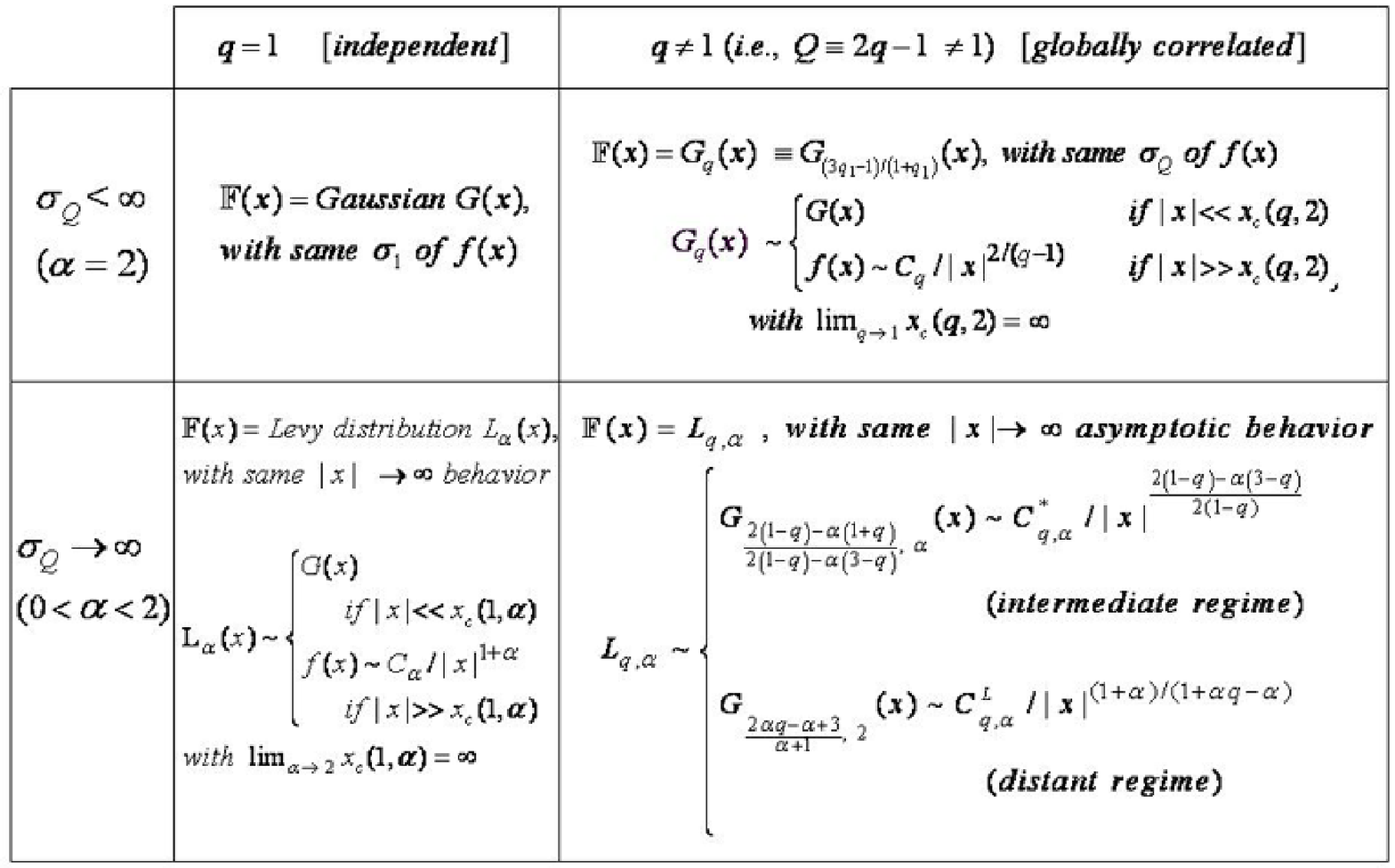}
\end{center}
\caption{\small $N^{1/[\alpha(2-q)]}$-scaled attractors ${\cal F}(x)$ when summing $N \to\infty$ 
$q$-independent 
identical random variables with symmetric distribution $f(x)$ with $Q$-variance $\sigma_Q \equiv 
\int_{-\infty}^\infty dx \, x^2 \, [f(x)]^Q /\int_{-\infty}^\infty dx \, [f(x)]^Q $ $(Q\equiv 2q-1; \, 
q_1=(1+q)/(3-q); \, q \ge 1)$. {\it Top left:} The attractor is the Gaussian sharing with $f(x)$ the same 
variance $\sigma_1$  (standard CLT). {\it Bottom left:} The attractor is the $\alpha$-stable L\'evy 
distribution which shares with $f(x)$ the same asymptotic behavior, i.e., the coefficient $C_\alpha$ 
(L\'evy-Gnedenko CLT, or $\alpha$-generalization of the standard CLT). {\it Top right:} The attractor is the 
$q$-Gaussian which shares with $f(x)$ the  same $(2q-1)$-variance, i.e., the coefficient $C_q$ 
($q$-generalization of the standard CLT, or $q$-CLT). {\it Bottom right:} The attractor is the 
$(q,\alpha)$-stable distribution which shares with $f(x)$ the  same asymptotic behavior, i.e., the coefficient 
$C^L_{q,\alpha}$ ($q$-generalization of the L\'evy-Gnedenko CLT and $\alpha$-generalization of the $q$-CLT). 
The case $\alpha < 2$, for both $q=1$ and $q \ne 1$ (more precisely $q>1$), further demands specific 
asymptotics for the attractors to be those indicated; essentially the divergent $q$-variance must be due to 
fat tails of the power-law class, excepting for possible logarithmic corrections (for the $q=1$ case see, 
for instance, \cite{BouchaudGeorges1990} and references therein).   
}
\label{figtheorem}
\end{figure*}

The {\it Central Limit Theorem} within theory of probabilities basically states that the {\it sum} of a large 
number $N$ of {\it independent} (or quasi-independent in some specific sense) random variables whose variance 
is {\it finite} converge, after appropriate centering and rescaling, to a Gaussian 
(i.e., $p(x) \propto e^{- \beta \,x^2}$). This distribution constitutes an attractor in the space of 
distributions, and is therefore thought to be the reason for the ubiquity of Gaussian distributions in nature. 
If the single distribution has a {\it divergent} variance instead (and also satisfies some supplementary 
mathematical conditions), the attractors are the celebrated L\'evy distributions. The situation changes 
drastically if strong correlations exist among the $N$ random variables. Depending on the nature of the 
correlations very many types of attractors might emerge. There is however a special class of strong 
correlations, referred to as {\it $q$-independence} \cite{UmarovTsallisSteinberg2008}, for which the 
attractors are $q$-Gaussians (i.e., $p(x) \propto e_q^{-\beta\,x^2}$, where the $q$-exponential function is 
defined as the inverse of the $q$-logarithmic one defined in Eq. (\ref{qlogarithmic})), if a specific 
generalized variance is {\it finite}. If this variance {\it diverges} instead, the attractors are the 
so-called $(q,\alpha)$-stable distributions: see \cite{UmarovTsallisGellMannSteinberg2009} for full details. 
The schematic description of these four theorems is presented in Fig. \ref{figtheorem}.

The physical-mathematical interpretation of the class of strong correlations named as $q$-independence is not 
yet fully ellucidated. However, it might well be that $q$-independence between $N$ random variables implies 
(strict or asymptotic) probabilistic scale-invariance in the sense that
\begin{equation}
\int dx_N\,h_N(x_1,x_2,...,x_N) \sim h_{N-1}(x_1,x_2,...,x_{N-1}) \;\;\;(N\to\infty) \,. 
\label{scaleinvariance}
\end{equation}
Although probably necessary, this property is surely not sufficient. Indeed, (strictly or asymptotically) scale-invariant probabilistic models (with finite values for the appropriately generalized variance) have been 
analytically solved, some of them yielding, in the $N\to\infty$ limit, $q$-Gaussians, whereas other models 
yield distributions which numerically are amazingly close to $q$-Gaussians, but which definitively are 
{\it not} exactly $q$-Gaussians. 
Models that yield $q$-Gaussians are available in \cite{RodriguezSchwammleTsallis2008,HanelThurnerTsallis2009}; 
models that have been proved \cite{HilhorstSchehr2007} to be not exactly $q$-Gaussians are presented in 
\cite{MoyanoTsallisGellMann2006,ThistletonMarshNelsonTsallis2009}.

Like Gaussians, $q$-Gaussians also are ubiquitous \footnote{The word {\it ubiquitous} is here used not in the 
strict sense of being everywhere, but only in the loose sense of being found very frequently.} in natural, 
artificial and social systems. What could be the cause of such fact? It could very well be precisely the 
theorem appearing in Fig. \ref{figtheorem} which corresponds to $q$-Gaussian attractors. In what follows we 
shall exhibit various nearly $q$-Gaussian  distributions: in Section 2 for dissipative 
one-dimensional dissipative maps, in Section 3 for long-range-interacting many-body classical Hamiltonians.
These two systems share a crucial property, namely that they have a maximal Lyapunov exponent which approaches zero, thus excluding strong chaos.
Finite-size or finite-precision 
effects are present in them: we mimic this property in Section 4 with a simple mathematical 
model.   
Finally we conclude in Section 5 by mentionning various systems presented in the literature which also appear to exhibit $q$-Gaussians. 

\section{Unimodal one-dimensional dissipative maps}

Let us here concentrate on a paradigmatic one-dimensional dissipative dynamical system, namely the logistic 
map, defined as 
$x_{t+1}=1-a x_t^2$, where $a$ is the map parameter ($0\le a \le2$), $-1 \le x_t \le1$, and $t=0,1,2,...$. 
Our object of interest is the sum 

\begin{equation}
y= \sum_{i=N_0+1}^{N_0+N} (x_i -\langle x \rangle)
\label{y}
\end{equation} 
in the vicinity of chaos threshold $a_c=1.4011..$, where $N_0$ is the number of transient steps 
(typically $N_0 >>1$) and 
\begin{equation}
\langle x \rangle = \frac{1}{n_{ini}} \frac{1}{N} \sum_{j=1}^{n_{ini}} \sum_{i=1}^N x_i^{(j)}\; 
\end{equation}
is the average over a large number of $N$ iterates as well as a large number $n_{ini}$ of randomly chosen 
initial values $x_1^{(j)}$ of iterates of the map. 
This problem has been addressed firstly in \cite{TirnakliBeckTsallis2007} and a closer look has been given 
very recently in \cite{TirnakliTsallisBeck2009}. Here, we try to further clarify the study and analyse it 
in a more compact manner.

Generically, the problem at hand is the form of the probability distribution of the random variable given 
in Eq.~(\ref{y}). Indeed, the ordinary Central Limit Theorem (CLT) (yielding the Gaussian form) is applicable 
when the Lyapunov exponent is positive, but, when approaching the edge of chaos from above 
(i.e., for $a>a_c$), an infinite number of values of $a$ accumulate which violate this condition. 
The situation becomes then quite subtle, as we shall review here. Essentially, strong correlations between 
the iterates of the map emerge. In \cite{TirnakliTsallisBeck2009} it is  shown that the problem is much 
more complex than the ordinary CLT, which is the case when the map is at the chaotic regime (e.g., for $a=2$). 
One needs now to be careful on how we approach the chaos threshold point ($a_c$) and how the number of 
iterations ($N$) to be used increases to infinity. In mathematical language, this means that two limits are 
to  be performed simultaneously, namely, $(a-a_c) \to 0$ and $1/N\to 0$. 
It is argued in \cite{TirnakliTsallisBeck2009} that the limit distributions appear to be of $q$-Gaussian 
type if these two limits are performed simultaneously in the following special way (which satisfies an 
appropriate scaling relation). We first choose a value of $a$ above and close to $a_c$. 
We then calculate the quantity $n$ defined as follows:
\begin{equation}
n= - \frac{\log |a-a_c|}{\log \delta} \,,
\label{scaling}
\end{equation}
where $\delta =4.6692011...$ is the Feigenbaum constant. We then denote by $k$ the nearest integer value of 
$2n$, and define $N^*$ through
\begin{equation}
N^*= 2^{k} \,.
\end{equation}
The $q$-Gaussians numerically appear to gradually emerge when we choose $N=N^*$ and keep making 
$(a-a_c) \to 0$ (hence $N =N^* \to\infty$).  
Let us refer to this region as the $q$-Gaussian probability distribution functions (PDFs) one and analyse 
its borders. 
To do this, typical values for $a$ have been used: they are given in Table~1, as well as their 
related parameters; $a$ values can easily be taken so that $2n$ values would be obtained with the same 
precision. Each group with the same precision enables the construction of a linear curve in the space 
$1/N$ vs $(a-a_c)^s$, where $s \equiv \ln 4 /  \ln \delta \simeq 0.9$: see Fig.~3. 
Numerical inspection has shown that no other $q$-Gaussian linear curves occur at the left of the 
largest slope and at the right of the smallest slope in Fig. 3.
All $q$-Gaussian lines appear to exist only between these two extremes; see examples 
in Fig~\ref{q-gauss} as well as Fig.~\ref{qbeta} where $(q,\beta)$ pairs of all studied cases are plotted.

\begin{table}[h]
\begin{center}
\caption{The values of map parameter $a$ and $N^*$ used in this work. The values of $2n$  (obtained from 
the scaling relation), and the values of $q$ and $\beta$ (estimated from simulations) are also listed. 
In this work, the critical value $a_c$  is approximated as $a_c=1.4011551890920505$.} 
\vspace{0.6cm}
\begin{tabular}{|l|l|l|l|l|l|}
\hline \hline
 $a$            & $a-a_c$    &   $2n$   & $N^*=2^{k}$ & $q$   & $\beta$ \\
\hline \hline
 1.40159888     & 0.00044369 & 10.02    & $2^{10}$ &          &        \\ \cline{1-3}
 1.40125021721  & 0.00009503 & 12.02    & $2^{12}$ & 1.70     &  6.6    \\ \cline{1-3}
 1.40117554121  & 0.00002035 & 14.02    & $2^{14}$ &          &         \\ \cline{1-3}
 1.40115954790  & 0.00000436 & 16.02    & $2^{16}$ &          &         \\ \cline{1-3}
\hline
 1.40152683     & 0.00037164 & 10.25    & $2^{10}$ &          &         \\ \cline{1-3}
 1.40123478     & 0.00007959 & 12.25    & $2^{12}$ & 1.70     &  6.5    \\ \cline{1-3}
 1.401172235    & 0.00001705 & 14.25    & $2^{14}$ &          &         \\ \cline{1-3}
 1.4011588398   & 0.00000365 & 16.25    & $2^{16}$ &          &         \\ \cline{1-3}
\hline
 1.4014862      & 0.00033101 & 10.40    & $2^{10}$ &          &         \\ \cline{1-3}
 1.401226075    & 0.00007088 & 12.40    & $2^{12}$ & 1.70     &  6.8    \\ \cline{1-3}
 1.401170372    & 0.00001518 & 14.40    & $2^{14}$ &          &         \\ \cline{1-3}
 1.401158441    & 0.00000325 & 16.40    & $2^{16}$ &          &         \\ \cline{1-3}
\hline
 1.401464065    & 0.00030888 & 10.49    & $2^{10}$ &          &         \\ \cline{1-3}
 1.401221341    & 0.00006615 & 12.49    & $2^{12}$ & 1.68     &  6.7    \\ \cline{1-3}
 1.4011693567   & 0.00001417 & 14.49    & $2^{14}$ &          &         \\ \cline{1-3}
 1.4011582234   & 0.00003034 & 16.49    & $2^{16}$ &          &         \\ \cline{1-3}
\hline
1.40145934      & 0.00030415 & 10.51    & $2^{11}$ &          &         \\ \cline{1-3}
 1.40122033     & 0.00006514 & 12.51    & $2^{13}$ & 1.61     &  6.8    \\ \cline{1-3}
 1.40116914     & 0.00001395 & 14.51    & $2^{15}$ &          &         \\ \cline{1-3}
 1.401158177    & 0.00000299 & 16.51    & $2^{17}$ &          &         \\ \cline{1-3}
\hline
1.40138924      & 0.00023405 & 10.85    & $2^{11}$ &          &         \\ \cline{1-3}
 1.401205317    & 0.00005013 & 12.85    & $2^{13}$ & 1.65     &  6.5    \\ \cline{1-3}
 1.401165925    & 0.00001074 & 14.85    & $2^{15}$ &          &         \\ \cline{1-3}
 1.4011574883   & 0.00000230 & 16.85    & $2^{17}$ &          &         \\ \cline{1-3}
\hline
1.40136531      & 0.00021012 & 10.99    & $2^{11}$ &          &         \\ \cline{1-3}
 1.40120019     & 0.00004500 & 12.99    & $2^{13}$ & 1.63     &  6.5    \\ \cline{1-3}
 1.401164827    & 0.00000964 & 14.99    & $2^{15}$ &          &         \\ \cline{1-3}
 1.4011572532   & 0.00000206 & 16.99    & $2^{17}$ &          &         \\ \cline{1-3}
\hline
\end{tabular}
\end{center}
\label{avalues}
\end{table}

\begin{figure}
\begin{center}
\includegraphics[height=10cm]{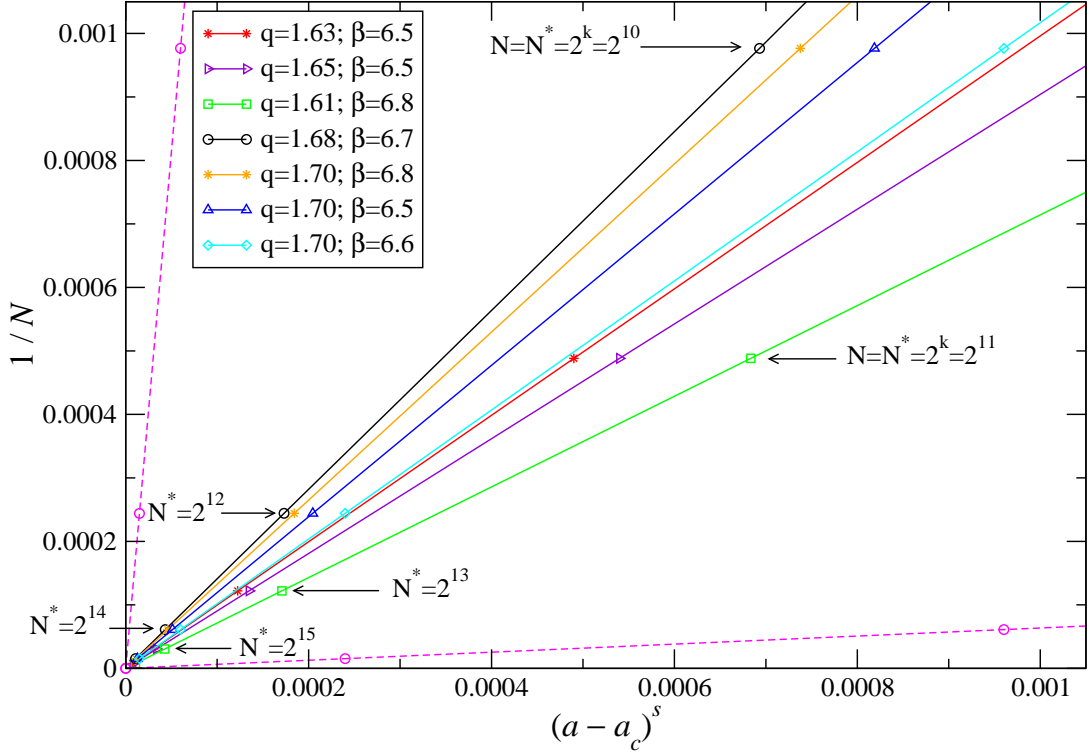}
\caption{View of the parameter region $1/N$ vs $(a-a_c)^s$ with $s=\ln 4 / \ln \delta\simeq0.9$ coming 
directly from the scaling relation (\ref{scaling}). 
Typical values of $a$ are shown, and their correspondiong values of $N^*$ are determined from the scaling 
relation (\ref{scaling}). 
Any other possible choices for $N^*$ yield lines that remain between the lines with the largest 
and the smallest slopes shown in the figure. If one approaches the critical point (origin) along any 
of these lines in this region, the probability distribution function (PDF) appears to gradually approach, 
excepting for a small oscillating contribution, a $q$-Gaussian. The regions at the left of the largest slope 
and at the right of the smallest slope are not accessible as far as $N^*$ values are concerned. Four of the 
seven $q$-Gaussian examples are presented in Fig.~\ref{q-gauss}; the peaked (almost vertical line) 
and the Gaussian (almost horizontal line) examples are presented in Fig.~\ref{regions}.}
\end{center}
\label{lines}
\end{figure}

\begin{figure}[h]
\begin{minipage}{18pc}
\includegraphics[width=18pc]{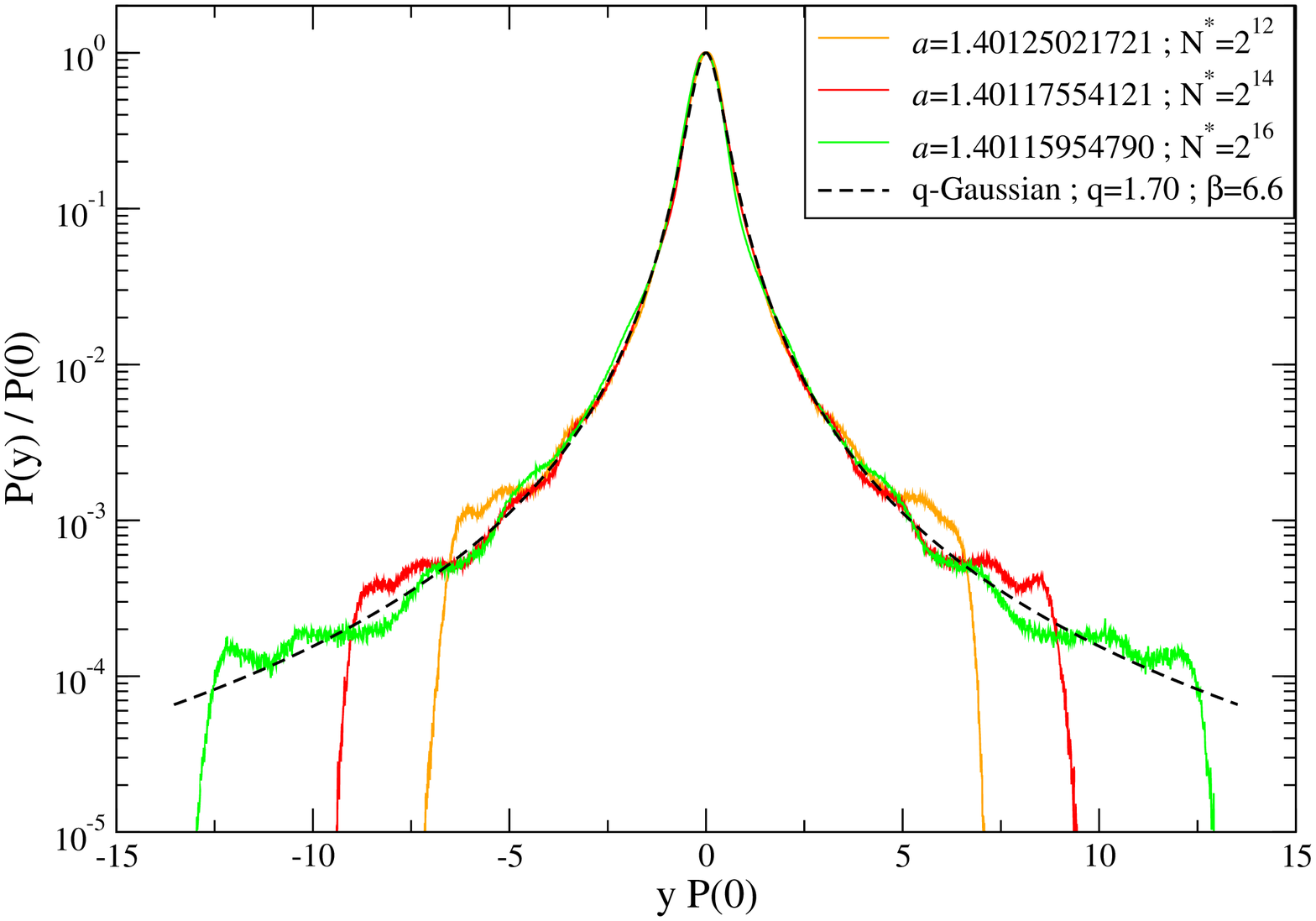}
\end{minipage}
\hspace{2pc}%
\begin{minipage}{18pc}
\includegraphics[width=18pc]{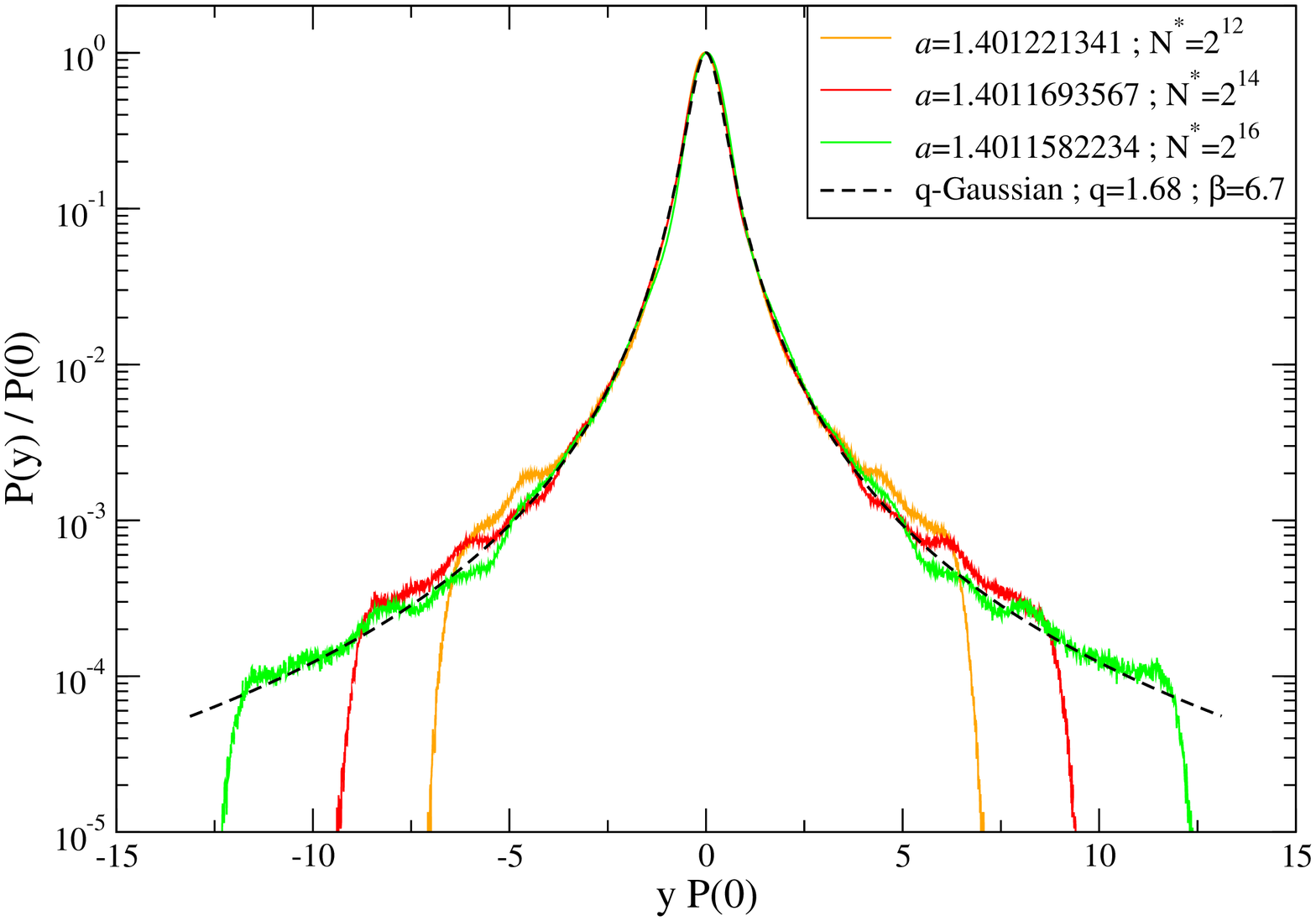}
\end{minipage} \\
\hspace{2pc}
\begin{minipage}{18pc}
\vspace{1.0cm}
\includegraphics[width=18pc]{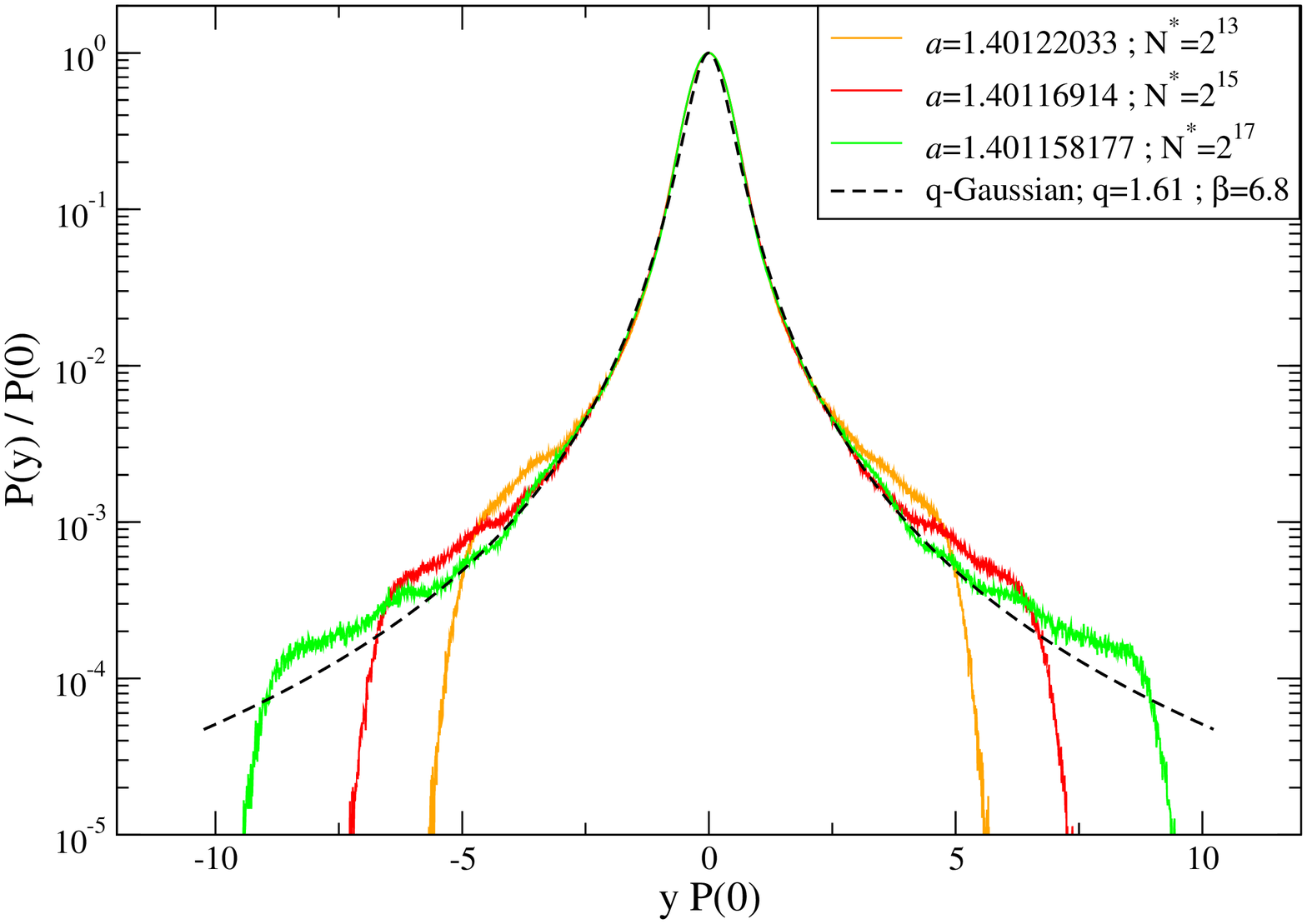}
\end{minipage}
\hspace{2pc}%
\begin{minipage}{18pc}
\vspace{1.0cm}
\includegraphics[width=18pc]{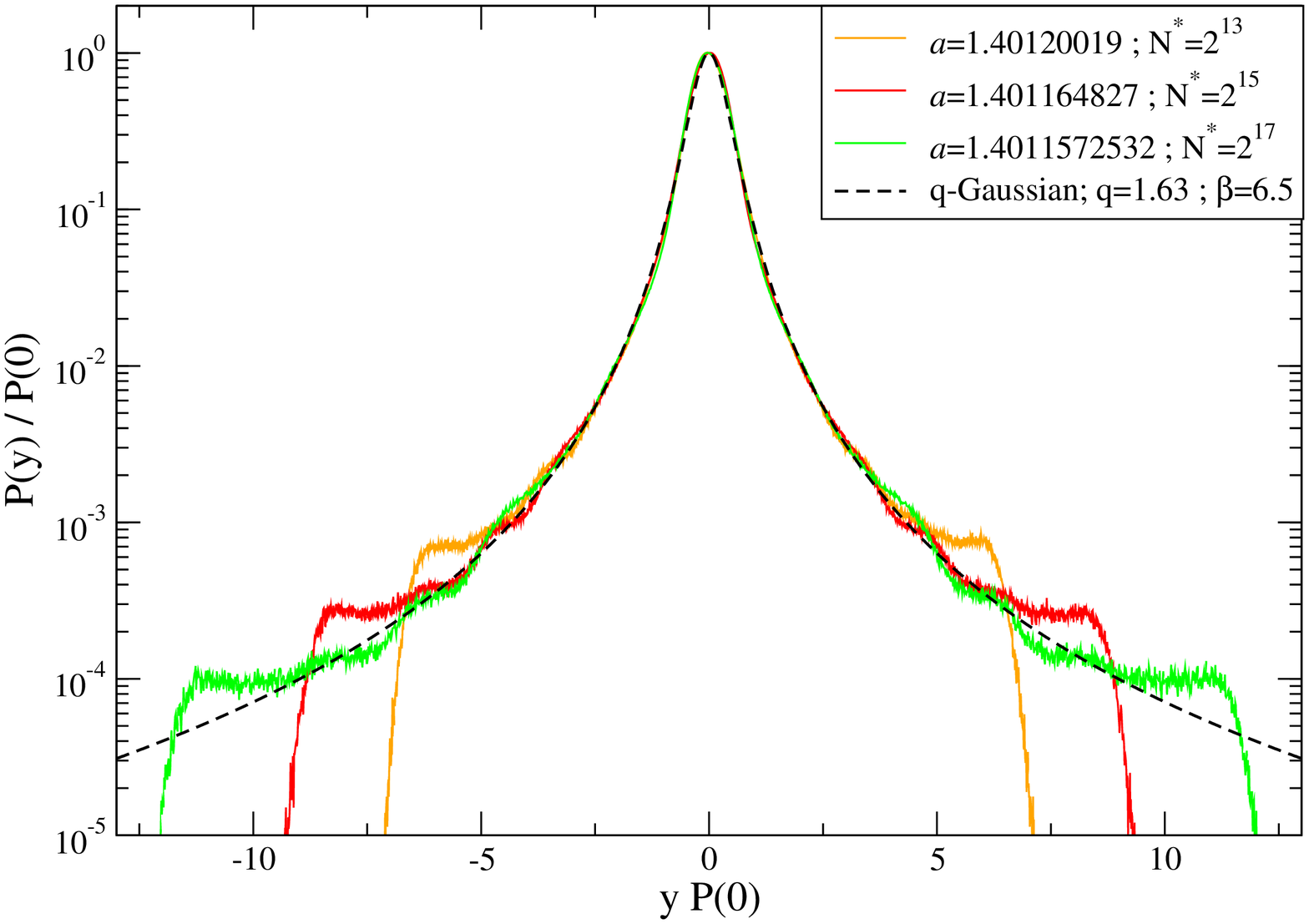}
\end{minipage}
\hspace{2pc}%
\caption{Data collapse of PDF's for the cases $N=N^*=2^{k}$ for four of the seven representative 
examples given in Table~1. The (possible) $q$-Gaussians are approached through finite-$N$ 
effects analogous to those exhibited in Fig.~\ref{crossover2}.}
\label{q-gauss}
\end{figure}

\begin{figure}[h]
\includegraphics[width=28pc]{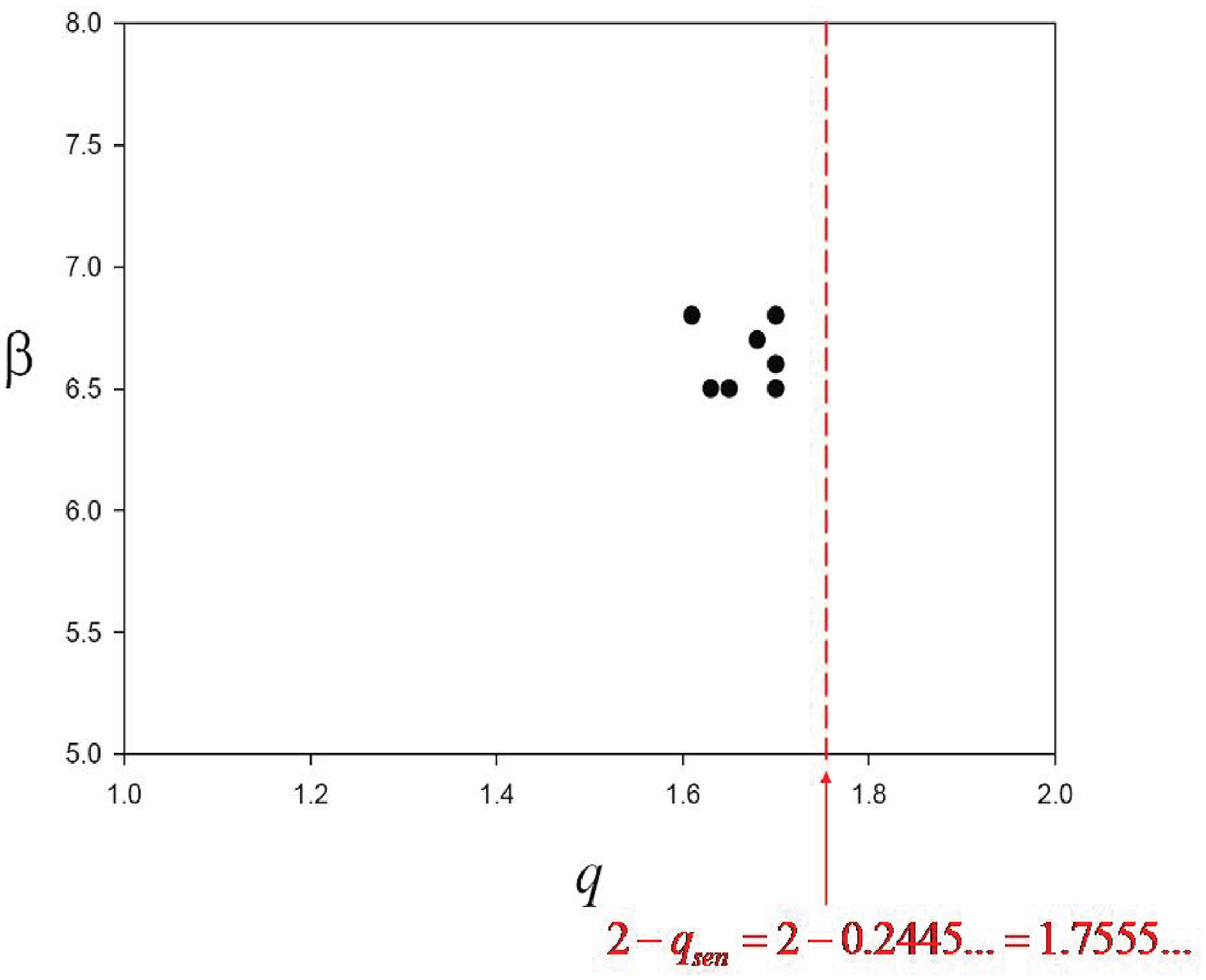}
\caption{The parameters $(q,\beta)$ corresponding to the seven $q$-Gaussians indicated in Fig. 3.
These specific seven examples appear to exclude the value $2-q_{sen}=1.7555...$, which could have been a 
plausible result. At the present numerical precision, even if quite high, it is not possible to infer whether 
the analytical result corresponding to the present observations would be only one or  a set of $q$-Gaussians, 
assuming that exact $q$-Gaussians are involved, on top of which a small oscillating component possibly exists.}
\label{qbeta}
\end{figure}

The $q$-Gaussian-like PDFs are lost as soon as the scaling 
relation (Eq.~(\ref{scaling})) is ignored, i.e., if values for $N$ larger or smaller that $N^*$ are used. 
Two new regions emerge. If we use values for $N$ that are sensibly smaller than $N^*$, or, in other words, 
the value of $a$ that is being used is too close to the critical point, peaked PDF's are observed (see also 
\cite{RobledoFuentes2009}). On the other extreme, if we use values for $N$ that are sensibly larger than 
$N^*$, we observe in most of the cases Gaussian PDF's, i.e., $P(y)=e^{-y^2/(2\sigma^2)}/\sqrt{2\pi \sigma^2}$). Two representative examples for the two regions outside the $q$-Gaussian one are indicated 
with dashed lines (magenta) in Fig.~3 and illustrated in Fig.~6.

\begin{figure}[h]
\vspace{2.0cm}
\begin{minipage}{18pc}
\includegraphics[width=18pc]{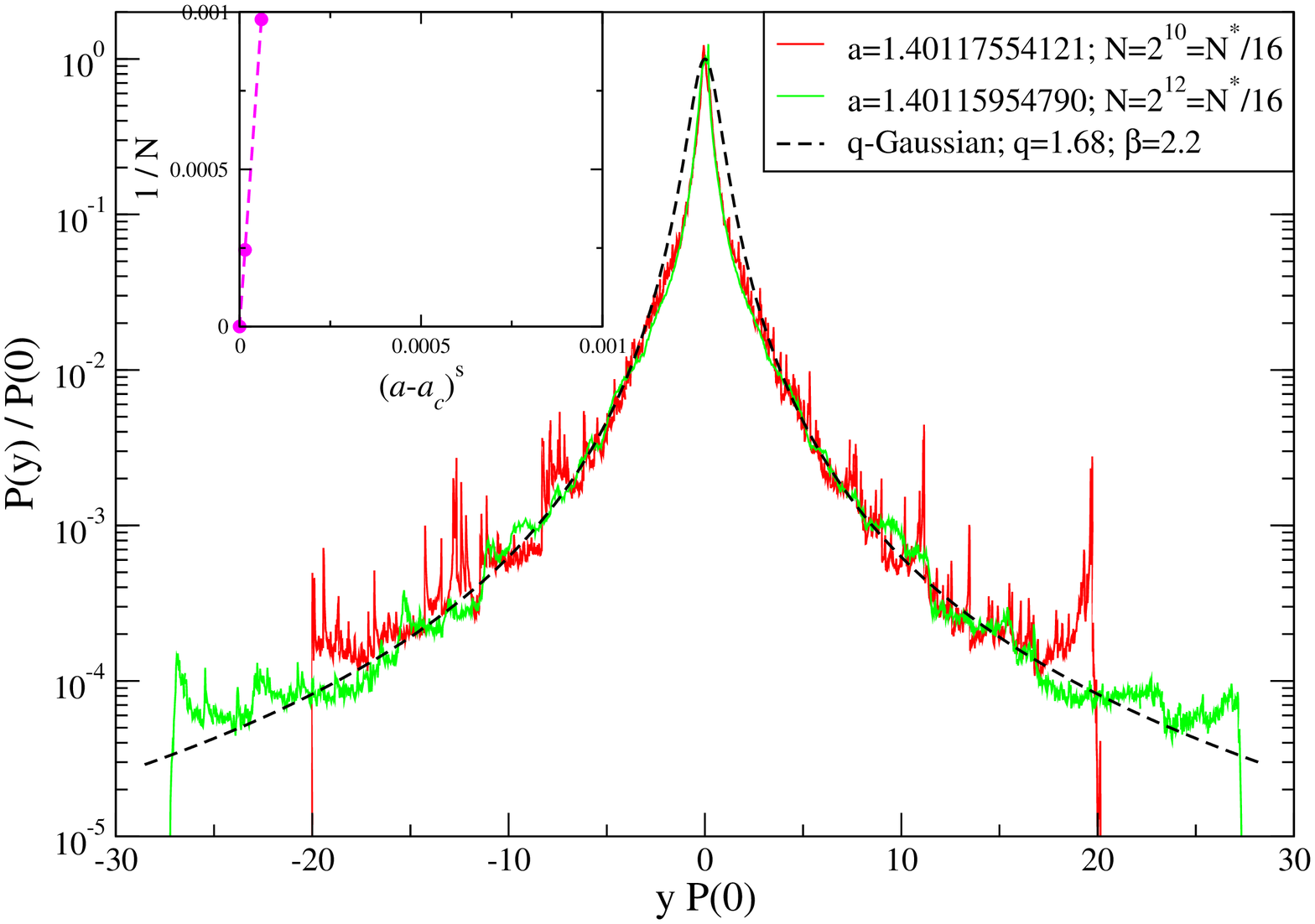}
\end{minipage}
\hspace{2pc}%
\begin{minipage}{18pc}
\includegraphics[width=18pc]{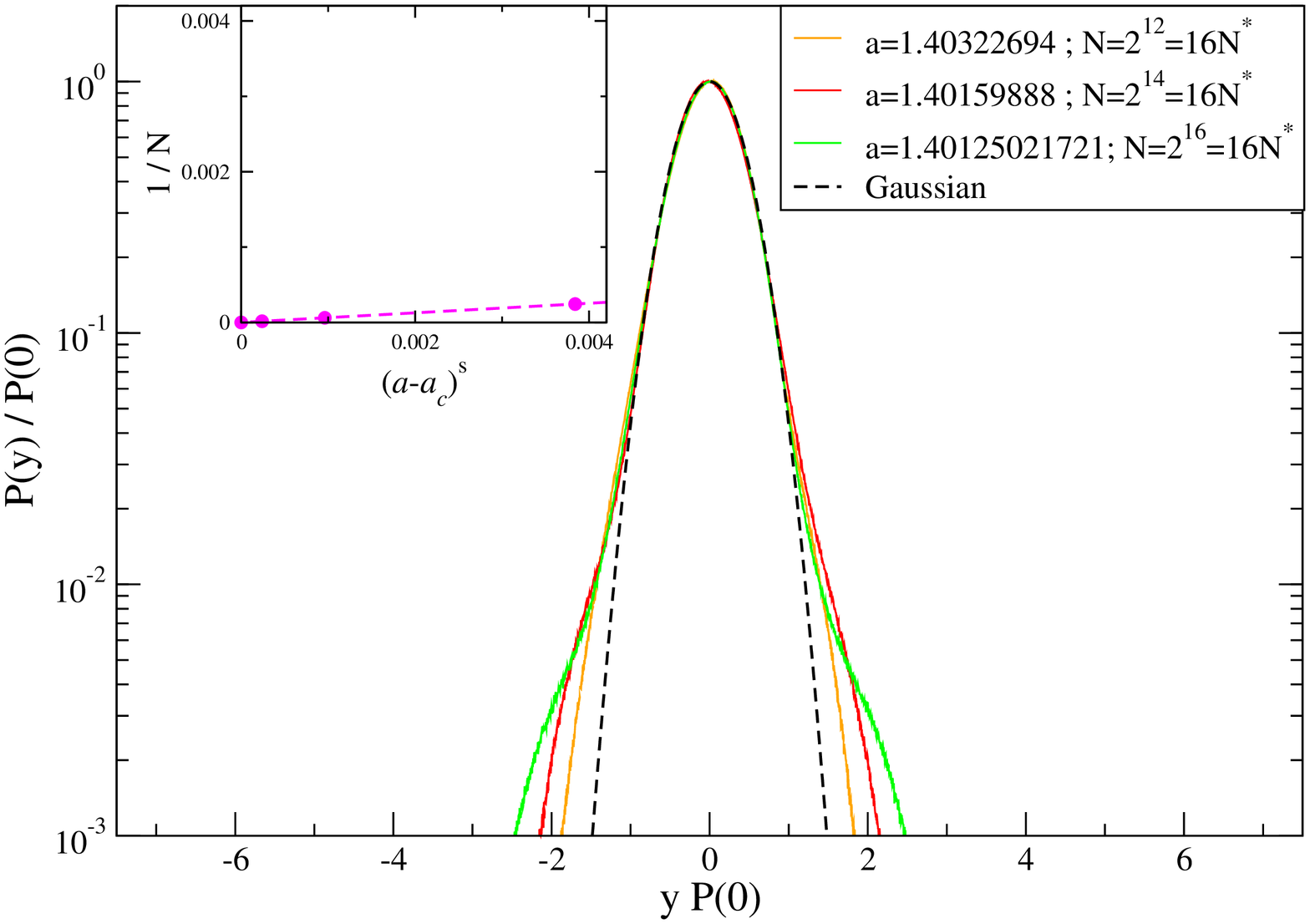}
\end{minipage}
\hspace{2pc}%
\caption{Representative examples from the peaked region (left) and the Gaussian region (right).}
\label{regions}
\end{figure}

\section{Long-range-interacting many-body classical Hamiltonian systems}

\begin{figure}[h]
\includegraphics[width=30pc]{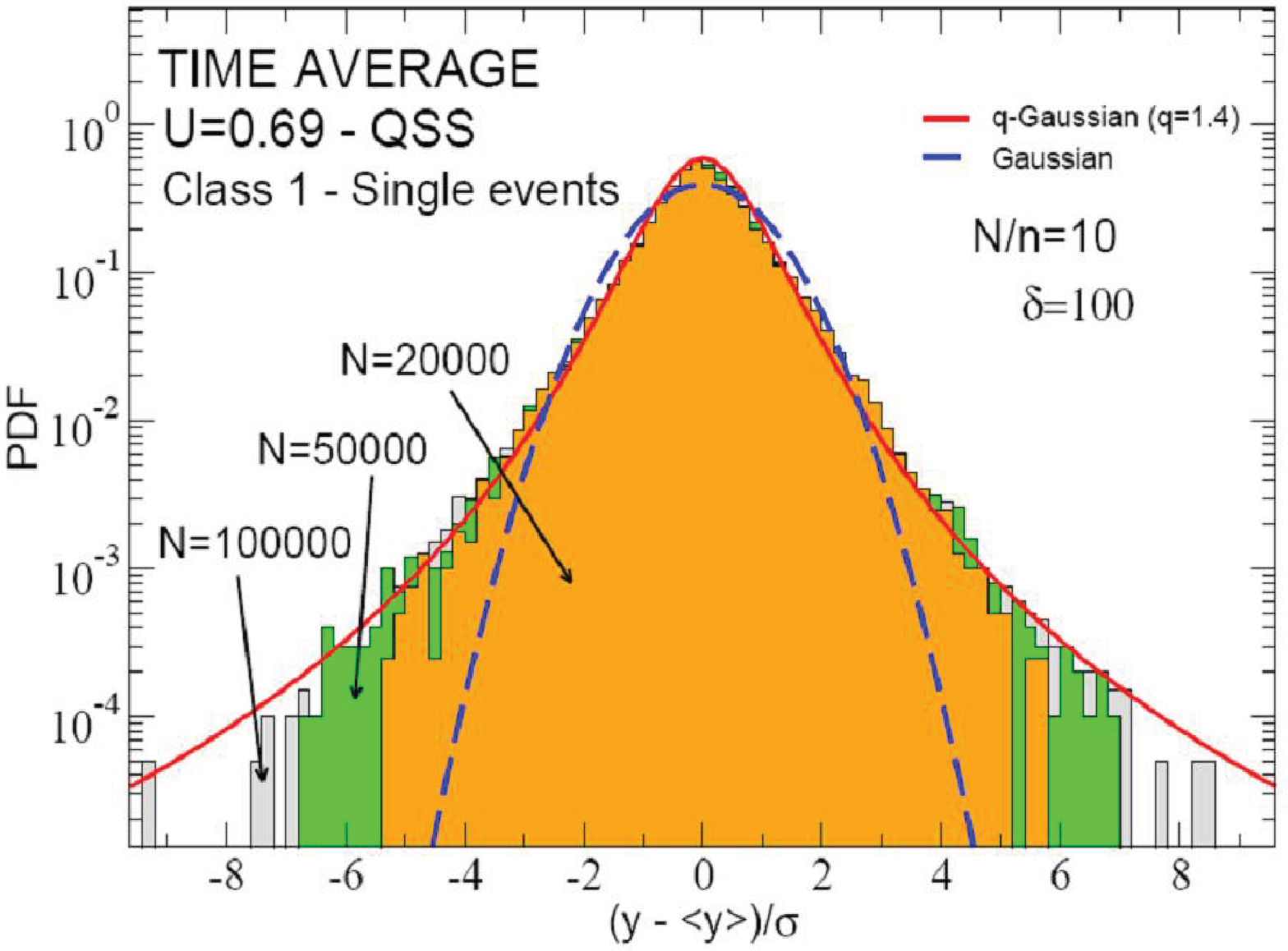}
\caption{Velocity distribution for the HMF model at its quasi-stationary state (QSS) [Data from Fig.~2 of 
\cite{PluchinoRapisardaTsallis2009}]. The finite-$N$ effects are visible, gradually approaching a (possible) 
$q$-Gaussian when both $N$ and $n$ are diverging. An analogous behavior is shown in Fig.~\ref{crossover2}. For comparison, a Gaussian is shown as well.}
\label{HMF}
\end{figure}

Let us consider the following classical Hamiltonian of $N$ interacting planar classical rotators on a 
$d$-dimensional (simple hypercubic) lattice \cite{AnteneodoTsallis1998}:
\begin{equation}
{\cal H}=\sum_{i=1}^N \frac{p_i^2}{2}+\frac{1}{2}\sum_{i,j} \frac{1-cos(\theta_i-\theta_j)}{r_{ij}^\alpha}  
\;\;\;\;(\alpha \ge 0) \,,
\end{equation}
where $r_{ij}$ runs over all possible distances within the $d$-dimensional lattice. The particular case 
$\alpha=0$ is referred to in the literature as the $HMF$  model \cite{AntoniRuffo1995}, and has been 
intensively studied in the last decade (in its standard representation the coupling constant is divided by 
$N$, which artificially makes the total energy extensive in the thermodynamical sense). Its dynamical 
molecular approach has exhibited a variety of interesting phenomena: see, for instance, \cite{Tsallis2009a} 
and references therein. In particular, for the isolated system (microcanonical ensemble) at energy per 
particle equal to 0.69, long-standing quasi-stationary states ($QSS$) emerge when certain classes of initial 
conditions (usually called {\it water-bag} initial conditions) are used. Within the water-bag conditions, one 
may consider initial magnetization equal to zero (usually referred to as $M=0$), or equal to its maximal value 
(usually referred to as $M=1$), or values in between. Such choices influence the specific trajectory of the 
full system within its $2N$-dimensional phase space (Gibbs $\Gamma$ space). It has been shown that, for $M=1$ 
initial conditions (possibly for virtually all values of initial $M$), {\it ergodicity is broken}. Indeed the 
{\it summed} (over $n$ equidistant instants) one-velocity marginal PDF differs when we take ensemble-average or 
time-average: see \cite{PluchinoRapisardaTsallis2009} and references therein. Many of the time-averaged PDF's 
numerically approach a $q$-Gaussian (see Fig.~\ref{HMF} for one such example). Although no analytical proof is 
available at the present time, this might be a consequence of the $q$-Central Limit Theorem 
\cite{UmarovTsallisSteinberg2008}, within which $q$-Gaussians are the attractors in the space of PDF's. The 
finite-size effects are illustrated in Fig.~\ref{HMF}.

\section{Simple mathematical model for crossover between $q$-Gaussians}

As a mathematical simple model for finite-size or finite-precision effects exhibiting the gradual approach to 
$q$-Gaussians, let us consider the following differential equation \cite{Tsallis2009a,TsallisBemskiMendes1999}:
\begin{equation}
\frac{dy}{dx}=-a y^r - b y^q\;\;\;\;(b \ge a \ge0;\,q>r;\,y(0)=1)\,.
\label{diffeq}
\end{equation}
If $a=0$ the solution is
\begin{equation}
y=e_q^{-bx} \,.
\end{equation}
If $b=0$ the solution is
\begin{equation}
y=e_r^{-ax} \,.
\end{equation}
If $b>a>0$ a crossover occurs from the $q$-exponential solution for $x$ not too large to the $r$-exponential 
solution for $x$ large enough; {\it increasing size or increasing precision} for specific models (such as the 
logistic map at its edge of chaos, or the $HMF$ Hamiltonian model at its $QSS$ state) acts analogously to 
{\it decreasing values of $a$} towards the limit $a=0$, for fixed $b$. Let us address this interesting case. 
>From (\ref{diffeq}) we obtain
\begin{eqnarray}
x=-\int_1^y \frac{du}{a u^r + b u^q}=
\frac{r}{b (-1+q) (q-r)}-\frac{ \left(r+(q-r)\, _2F_1\left[\frac{-1+q}{q-r},1,1+\frac{-1+q}{q-r},-\frac{a}{b}\right]\right)}{b (-1+q) (q-r)} \nonumber \\
-\frac{r y^{1-q}}{b (-1+q) (q-r)}+\frac{y^{1-q} \left(r+(q-r)\, _2F_1\left[\frac{-1+q}{q-r},1,1+\frac{-1+q}{q-r},-\frac{a y^{-q+r}}{b}\right]\right)}{b (-1+q) (q-r)}
\end{eqnarray}
where $_2F_1$ is the hypergeometric function. In general, this function does not admit an {\it explicit} 
expression in the form $y(x)$. An exception is the $r=1$ case, which yields
\begin{equation}
y=\frac{1}{\Bigl[(\frac{b}{a}+1)e^{(q-1)\,a\,x}-\frac{b}{a}    \Bigr]^{\frac{1}{q-1}}}\, .
\end{equation}
The $r=0$ case must be handled through the explicit $x(y)$ form, namely
\begin{equation}
x=\frac{1}{a}\Bigl\{ \\ _2F_1\Bigl[\frac{1}{q},1,1+\frac{1}{q},-\frac{b}{a} \Bigr] -y \, _2F_1\Bigl[\frac{1}{q},1,1+\frac{1}{q},-\frac{b}{a} \,y^q \Bigr]   \Bigr\} \,.
\end{equation}

Let us address now the case of the $q$-Gaussians. Following the form of Eq. (\ref{diffeq}), we consider
\begin{equation}
\frac{dy}{d(x^2)}=-a_r y^r - (a_q-a_r) y^q\;\;\;\;(a_q \ge a_r \ge0;\,q>r;\,y(0)=1)\,.
\label{diffeq2}
\end{equation}
If $a_r=0$, or equivalently if $r=q$, the solution is given by the $q$-Gaussian $y=e_q^{-a_q\,x^2}$. 
If $a_r=a_q$, the solution is given by the $r$-Gaussian $y=e_r^{-a_r\,x^2}$. For the case $a_q>a_r>0$ and 
$q>r$, we obtain a crossover between these two solutions, the $|x|\to\infty$ asymptotic one being the 
$r$-Gaussian behavior.

\begin{figure}[h]
\begin{minipage}{18pc}
\includegraphics[width=18pc]{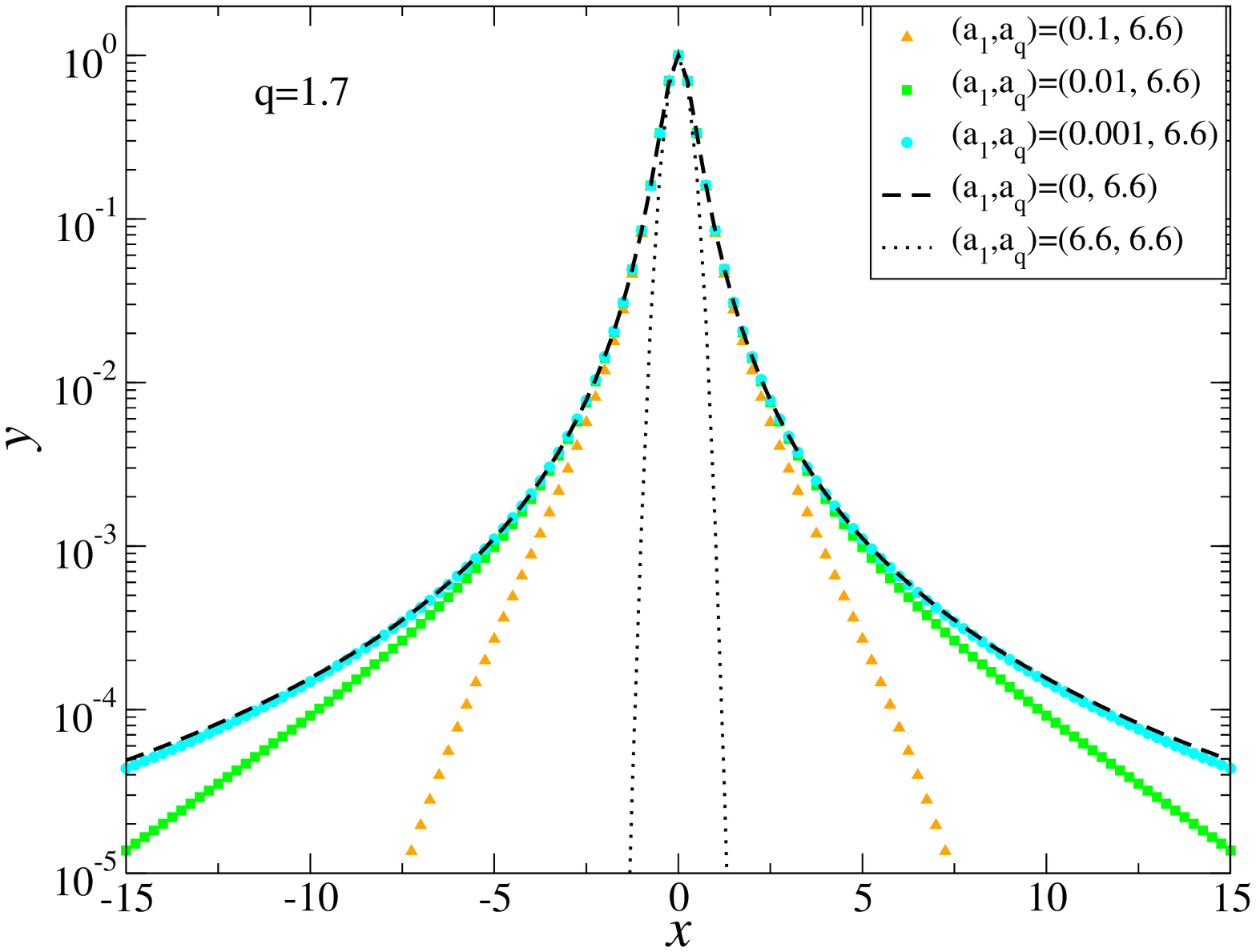}
\end{minipage}
\hspace{2pc}%
\begin{minipage}{18pc}
\includegraphics[width=18pc]{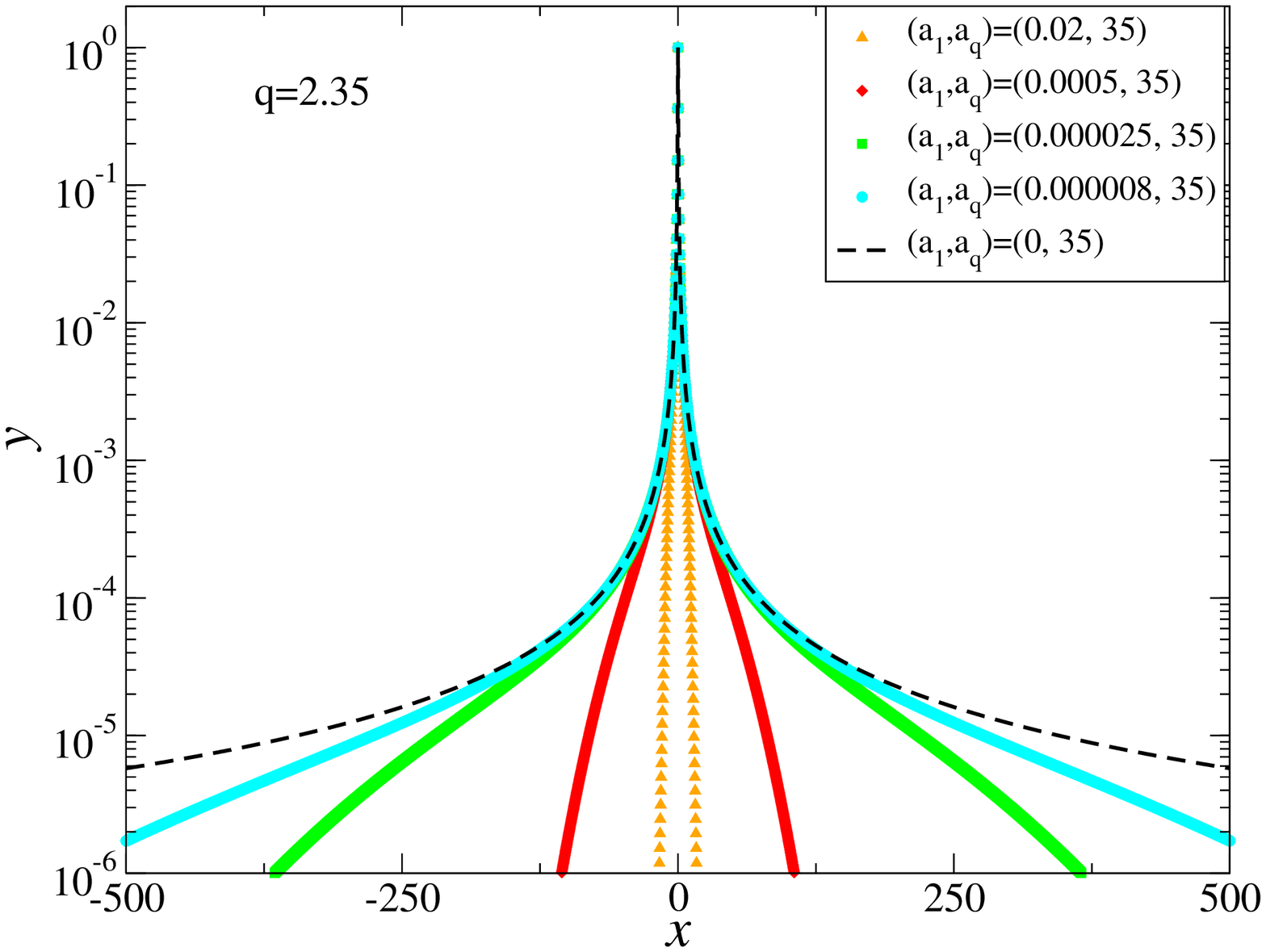}
\end{minipage}
\hspace{2pc}%
\caption{Crossover from $q$-Gaussian to successively distant Gaussians, i.e., illustrations of the case $r=1$ 
and $q>1$. {\it Left}: $q=1.7$; for comparison, a Gaussian is shown as well. {\it Right}: $q=2.35$.}
\label{crossover}
\end{figure}

\begin{figure}[h]
\vspace{1.5cm}
\includegraphics[width=30pc]{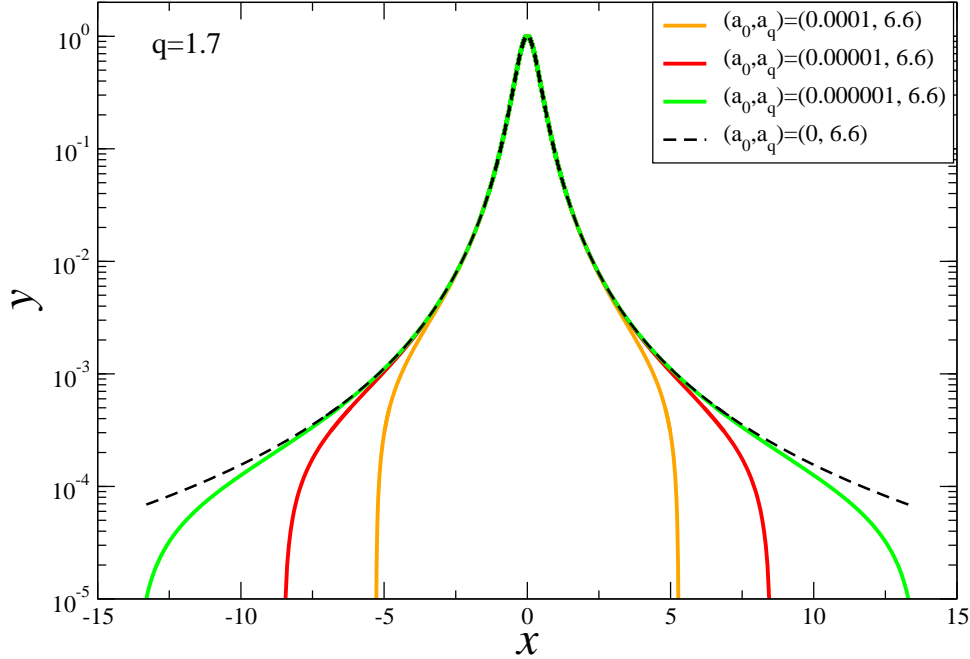}
\caption{Crossover from 1.7-Gaussian with $\beta=6.6$ to successively distant $r$-Gaussians with $r=0$.}
\label{crossover2}
\end{figure}

For $r=1$ and $q>1$, the solution is given by
\begin{equation}
y=\frac{1}{\Bigl[1-\frac{a_q}{a_1}+\frac{a_q}{a_1} \,e^{(q-1)a_1\,x^2}\Bigr]^{\frac{1}{q-1}}}\, .
\end{equation}
The general behavior of these solutions is given in Fig.~\ref{crossover}. It is evident from this 
figure that this solution is not the most appropriate one for the behavior observed in the neighborhood of the logistic map edge of chaos. 
We notice concomitantly that the appropriate solution for the logistic map seems to be very close to the one with $r=0$ and $q>1$, 
whose solution is given by 
\begin{equation}
x^2=\frac{1}{a}\Bigl\{ \\ _2F_1\Bigl[\frac{1}{q},1,1+\frac{1}{q},-\frac{b}{a} \Bigr] -y \, _2F_1\Bigl[\frac{1}{q},1,1+\frac{1}{q},-\frac{b}{a} \,y^q \Bigr]   \Bigr\} \,,
\end{equation}
with $a\equiv a_0$ and $b \equiv a_q-a_0$. Indeed, this solution seems to be a very good 
approximation for the behavior of PDFs obtained numerically for the logistic map. This can be seen 
immediately whenever the representative example of the solution given in Fig.~\ref{crossover2} is 
compared to the case given in Fig.~\ref{q-gauss}a.

\section{Final remarks}

Many $q$-Gaussian-like distributions have been observed in recent years in natural, artificial and social 
systems. Obviously, only for a mathematically formulated model, the hope exists to possibly prove analytically 
that the relevant distribution precisely, and not only approximatively, is a $q$-Gaussian. In all other cases, 
we can only expect for increasingly high-precision indications from real experiments or observations. 
Computational evidence can and does provide important hints, however never a proof.

This said, let mention in what follows some of the many other systems where $q$-Gaussians have been used to 
approach the observed PDF's: (i) The velocity distribution of (cells of) {\it Hydra viridissima} follows a 
$q=3/2$ PDF \cite{UpadhyayaRieuGlazierSawada2001}; (ii) The velocity distribution of (cells of) 
{\it Dictyostelium discoideum} follows a $q=5/3$ PDF in the vegetative state and a $q=2$ PDF in the starved 
state \cite{Reynolds2010}; (iii) The velocity distribution in defect turbulence \cite{DanielsBeckBodenschatz2004}; (iv)
The velocity distribution of cold atoms in a dissipative optical lattice 
\cite{DouglasBergaminiRenzoni2006}; (v) Velocity distribution during silo drainage \cite{ArevaloGarcimartinMaza2007a,ArevaloGarcimartinMaza2007b}; 
(vi) The velocity distribution in a driven-dissipative 2D dusty plasma, 
with $q=1.08\pm0.01$ and $q=1.05\pm 0.01$ at temperatures of $30000 \,K$ and $61000\, K$ respectively 
\cite{LiuGoree2008}; (vii) The spatial (Monte Carlo) distributions of a trapped $^{136}Ba^+$ ion cooled by various 
classical buffer gases at $300\,K$ \cite{DeVoe2009}; (viii) The distributions of price returns at the stock exchange 
\cite{Borland2002a,Borland2002b,Queiros2005}; (ix) The distributions of returns of magnetic field fluctuations 
in the solar wind plasma as observed in data from  Voyager 1 \cite{BurlagaVinas2005} and from Voyager 2 
\cite{BurlagaNess2009}; (x) The distributions of returns of the avalanche sizes in the Ehrenfest's dog-flea 
model \cite{BakarTirnakli2009}; (xi) The distributions of returns of the avalanche sizes in the self-organized 
critical Olami-Feder-Christensen model, as well as in real earthquakes 
\cite{CarusoPluchinoLatoraVinciguerraRapisarda2007}; (xii) The distributions of angles in the $HMF$ model 
\cite{MoyanoAnteneodo2006}; (xiii) The distribution of stellar rotational velocities in the Pleiades 
\cite{CarvalhoSilvaNascimentoMedeiros2008}. Some indirect evidence is available as well: although no $q$-Gaussian distribution has been directly observed in some relevant physical quantity, a $q$-exponential relaxation has been seen in various paradigmatic spin-glass substances through neutron spin echo experiments \cite{PickupCywinskiPappasFaragoFouquet2009}.

Clearly, the simplest hypothesis which would explain the ubiquity of $q$-Gaussians is the validity of the 
$q$-Central Limit theorem. The involved random variables would, in such case, be expected to be 
$q$-independent \cite{UmarovTsallisSteinberg2008}. The present belief is that (probabilistic) scale-invariance 
is necessary but not sufficient for $q$-independence. Further studies are needed to clarify the applicability 
of such ideas to the systems mentioned above, as well as possibly others.

\section*{Acknowledgments}
We thank insightful remarks by C. Beck, E.M.F. Curado, A. Pluchino, A. Rapisarda, A. Robledo and 
B.C.C. dos Santos. This work has been partially supported by CNPq and Faperj (Brazilian agencies), 
and by TUBITAK (Turkish agency) under the Research Project number 104T148.

\end{document}